\documentclass[]{aa}  

\usepackage{graphicx}
\usepackage{subfigure}
\usepackage{adjustbox}
\usepackage{hyperref}
\usepackage{txfonts}
\begin{document}

   \title{New ACV variables discovered in the \textit{Zwicky} Transient Facility survey}

   \author{B.~Bauer-Fasching\inst{1}
   \and K.~Bernhard\inst{2,3}
   \and E.~Br\"andli\inst{1}
   \and H.~Burger\inst{1}
   \and B.~Eisele\inst{1}
   \and S.~H{\"u}mmerich\inst{2,3}
   \and J.~Neuhold\inst{1}
   \and E.~Paunzen\inst{4}
   \and M.~Piecka\inst{1}
   \and S.~Ratzenb{\"o}ck\inst{1,5}
   \and M.~Pri{\v s}egen \inst{6}}

   \institute{Department of Astrophysics, Vienna University, T{\"u}rkenschanzstraße 17, 1180 Vienna, Austria
        \and Bundesdeutsche Arbeitsgemeinschaft f{\"u}r Ver{\"a}nderliche Sterne e.V. (BAV), Berlin,
              Germany 
          \and
              American Association of Variable Star Observers (AAVSO), Cambridge, USA
          \and Department of Theoretical Physics and Astrophysics, Faculty of Science, Masaryk University, Kotl\'{a}\v{r}sk\'{a} 2, 611 37 Brno, Czechia
   \and University of Vienna, Research Network Data Science at Uni Vienna, Kolingasse 14-16, 1090 Vienna, Austria
    \and Advanced Technologies Research Institute, Faculty of Materials Science and Technology in Trnava, Slovak University of Technology in Bratislava, Bottova 25, 917 24 Trnava, Slovakia}

\date{}
 
  \abstract
  {The manifestation of surface spots on magnetic chemically peculiar (mCP) stars is most commonly explained by the atomic diffusion theory, which requires a calm stellar atmosphere and only moderate rotation. While very successful and well described, this theory still needs to be revised and fine-tuned to the observations.}
  {Our study aims to enlarge the sample of known photometrically variable mCP stars (ACV variables) to pave the way for more robust and significant statistical studies. We derive accurate physical parameters for these objects and discuss our results in the framework of the atomic diffusion theory.}
  {We studied 1314 candidate ACV variables that were selected from the \textit{Zwicky} Transient Factory catalogue of periodic variables based on light curve characteristics. We investigated these objects using photometric criteria, a colour-magnitude diagram, and spectroscopic data from the LAMOST and \textit{Gaia} missions to confirm their status as ACV variables.}
  {We present a sample of 1232 new ACV variables, including information on distance from the Sun, mass, fractional age on the main sequence, fraction of the radius between the zero-age and terminal-age main sequence, and the equatorial velocity and its ratio to the critical velocity.}
  {Our results confirm that the employed selection process is highly efficient for detecting ACV variables. We have identified 38 stars with $v_\mathrm{equ}$ in excess of 150\,km\,s$^{-1}$ (with extreme values up to 260\,km\,s$^{-1}$). This challenges current theories that cannot explain the occurrence of such fast-rotating mCP stars.}


   \maketitle
%

\section{Introduction} \label{introduction} 

{The classical chemically peculiar (CP) stars of the upper main sequence 
have been targets of detailed investigations since their first detection by 
\citet{1897AnHar..28....1M}. They are excellent test objects for astrophysical processes 
such as diffusion, convection, and stratification in stellar atmospheres. 
There is a wide variety of peculiar stars in the spectral domain from 
B0 (30000\,K) to F5 (6500\,K). \citet{1974ARA&A..12..257P} divided the CP stars 
into four groups based on the presence of strong magnetic fields and
the kind of surface elemental peculiarity.}

{The magnetic chemically peculiar (mCP) stars, which encompass the Ap/CP2 stars and 
He-peculiar stars groups \citep{1974ARA&A..12..257P}, exhibit peculiar surface 
abundances and a non-uniform distribution of certain chemical elements. They
also possess strong and global magnetic fields. The CP2 stars show overabundances 
of Si, Sr, Cr, Eu, and the rare-earth elements as compared to the solar 
composition \citep[e.g.][]{1974ARA&A..12..257P,saffe2005,2009ssc..book.....G,ghazaryan18}. 
The He-peculiar stars comprise 
the B5 to B9 He-weak (CP4) stars that show anomalously weak \ion{He}{i} lines for
their temperature type, 
and the more massive B1 to B3 He-strong stars with anomalously strong \ion{He}{i} 
lines \citep{1998A&A...337..512A,ghazaryan19}. Other groups of CP stars include the metallic-line (Am/CP1) stars and the mercury-manganese (HgMn/CP3) 
stars; they, however, are only marginally relevant to the present study and will not 
be discussed here in detail.}

Many mCP stars exhibit strictly periodic changes in their spectra and brightness in different photometric 
passbands, which result from a flux redistribution in the unevenly distributed surface abundance patches (`chemical spots') of certain elements \citep{krticka2012} and are satisfactorily described by the oblique rotator model \citep{Stibbs1950}. Amplitudes of several hundredths up to some tenths of a magnitude are observed \citep{2017MNRAS.468.2745N}, and the amplitude of the light variability generally increases towards shorter wavelengths. These photometrically variable mCP stars are traditionally called $\alpha^{2}$ Canum Venaticorum (ACV) variables. Depending on the distribution of surface spots and the magnetic field characteristics, ACV variables show an amazing diversity of light curves \citep{2019A&A...622A.199J}.
{The latest version of the International Variable Star Index \citep[VSX;][]{2006SASS...25...47W} lists 1768 known ACV variables.}

For the manifestation of surface spots of different elements, a calm stellar atmosphere and only moderate rotation are required \citep{michaud1981}. Otherwise, the observed surface patterns become unstable and are destroyed. However, while the general framework of this complex mechanism has been well described, it still needs to be revised and adapted to observations. 

The present study aims to enlarge the sample of known ACV variables using data from the \textit{Zwicky} Transient Facility (ZTF) survey in order to pave the way for more robust and significant statistical studies. It follows up and concludes the feasibility study of \citet{2021A&A...656A.125F}, who identified 1400 candidate ACV variables via photometric criteria and published the 86 most promising candidates.

We present an investigation of the remaining 1314 candidates, taking into account, whenever available, spectroscopic data from the Large Sky Area Multi-Object Fiber Spectroscopic Telescope (LAMOST) and \textit{Gaia} missions to confirm their status as CP objects. We assembled a final sample of 1232 ACV variables, for which we derive astrophysical parameters and rotational velocities. We conclude by discussing our results in the framework of models that explain the CP star phenomenon.


Data sources are summarised in Section \ref{data_sources}. General methods and results are described in Section \ref{methods_and_results}; a search for membership in open star clusters is detailed in Section \ref{OCLs}. Finally, we discuss our results and conclude in Section \ref{conclusion}.

\begin{figure*}
        \includegraphics[width=2.1\columnwidth]{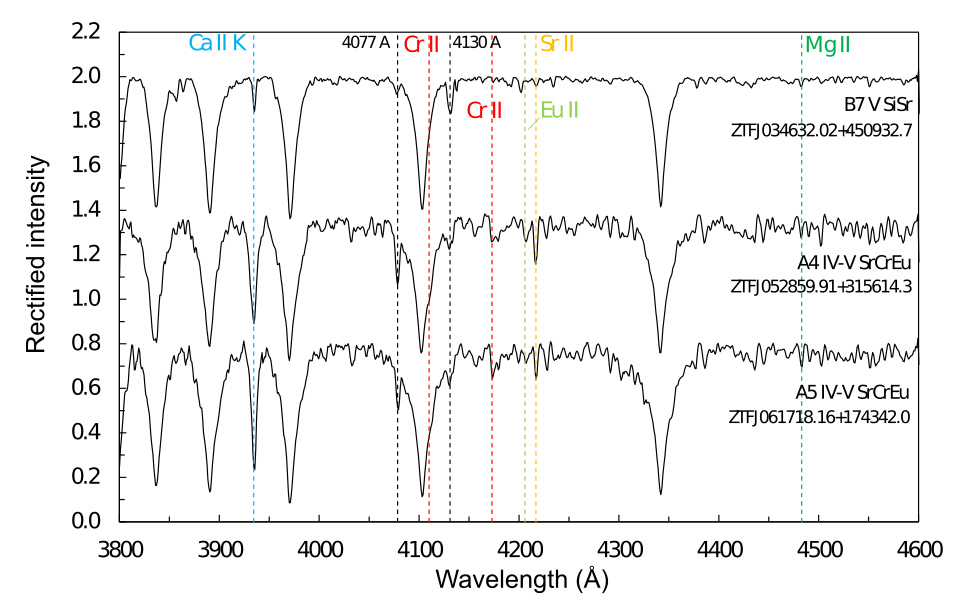}
    \caption{Blue-violet region of the LAMOST spectra of (from top to bottom) the CP2 stars ZTFJ034632.02+450932.7 (B7 V SiSr), ZTFJ052859.91+315614.3 (A4 IV-V SrCrEu), and ZTFJ061718.16+174342.0 (A5 IV-V SrCrEu). Some prominent lines of interest are identified.}
    \label{fig:showcase}
\end{figure*}

\section{Data sources}
\label{data_sources}

This section briefly summarises the employed data sources.

\subsection{The Zwicky Transient Facility survey} \label{ZTF}
In this study, we relied on data from the ZTF, a public-private partnership optical time-domain survey that uses the 48-inch Samuel 
Oschin Schmidt telescope situated at the Palomar Observatory in northern San Diego County, California, which is optimised for spectral classification of stars brighter than 19\,mag 
\citep{Graham_2019}, in conjunction with a wide field CCD camera with a 47 deg$^2$ field of 
view \citep{2019PASP..131a8002B}. In addition, an integral field unit spectrograph is employed on the Palomar 60-inch telescope that is used to identify transients over a significant fraction of the available sky and for dedicated spectroscopic follow-up observations. The acquired data were subsequently prepared in the {Infrared} Processing and Analysis Center (IPAC). The ZTF survey provides nearly 300 {epochs}
each year per object in the northern hemisphere and produces high-quality light curves in the $g$, $r$, and $i$ bands.

Due to the large area covered by the survey, the 
ZTF is a powerful facility for identifying transits or acquiring time-series photometric observations \citep{Graham_2019}. Its aims are the discovery of young supernovae and transits in general but also the study of active galactic nuclei, observation of asteroids and variable stars. For more information on the ZTF, we refer the reader to \citet{bellm19} and \citet{masci_2019}.

\subsection{The Large Sky Area Multi-Object Fibre Spectroscopic Telescope (LAMOST)} \label{LAMOST}

Whenever available, spectra from the LAMOST survey were employed in the investigation of our sample stars. LAMOST is operated at Xinglong Station Observatory in China and managed by the National Astronomical Observatories from the Chinese Academy of Sciences (NAOC). It {obtains} spectra from the northern sky using a reflecting Schmidt telescope with {primary mirror size of 4 metres} and 4000 movable fibres to survey an equal number of objects simultaneously.

We used LAMOST low-resolution spectra, which have a spectral resolution R\,$\sim$\,1800 and wavelength coverage from 3700\,\AA\ to 9000\,\AA. More information on the LAMOST project can be gleaned from \citet{lamost1} and \citet{lamost2}.

\subsection{The \textit{Gaia} mission BP/RP spectra} \label{BP_RP_spectra}

The \textit{Gaia} mission {\citep{2021A&A...649A...1G}} provides a wealth of low-resolution blue photometer and red photometer (BP/RP) spectra, which cover the wavelength region from 3300\,\AA\ to 10500\,\AA\ with a resolving power between 25 and 100, depending on the wavelength \citep{carrasco21}. Taking into account variations over the focal plane, all spectra are brought onto a common flux and pixel (pseudo-wavelength) scale. A mean spectrum is produced from multiple observations of the same source. \textit{Gaia} Data Release (DR) 3 already includes about 219\,000\,000 mean BP/RP spectra for objects up to the 18th magnitude in $G$. The signal-to-noise ratio depends on the apparent
magnitude and colour of the object. For more information on the \textit{Gaia} BP/RP spectra, we refer the reader to \citet{carrasco21} and \citet{montegriffo23}.

\section{Methods and results} \label{methods_and_results}

This section describes the methods and, where appropriate, the corresponding results obtained in this study.

\subsection{Target selection} \label{target_selection}

The present paper is a follow-on work to the feasibility study of \citet{2021A&A...656A.125F} and examines the 1314 candidate ACV variables that were identified but not further investigated by these authors. In the following, we provide a brief overview over the methodology employed by \citet{2021A&A...656A.125F} to select these stars.

Based on ZTF DR2, \citet{2020ApJS..249...18C} compiled a catalogue of 781\,602 periodic variable stars, which has been the principal source for target selection in \citet{2021A&A...656A.125F}. As ACV variables were not considered in the catalogue of \citet{2020ApJS..249...18C}, it does not have a category for this kind of object, and it can be assumed (and has indeed been proved by \citealt{2021A&A...656A.125F}) that ACV variables were consequently assigned to other categories, in particular to the class of the RS Canum Venaticorum (RS CVn) stars. These objects show rotational modulated light changes that, at least at first glance, resemble those of ACV variables. They are, however, close binary stars with active chromospheres and enhanced spot activity and thus a very different group of objects \citep[e.g.][]{hall76}, which can easily be distinguished from ACV variables by, for example, the use of a colour-magnitude diagram (CMD).
{The newest data of the \textit{Gaia} mission now allow us to establish the CMD 
precisely and, therefore, to further select between ACV and RS CVn stars.}

To identify ACV candidates among the 81\,393 RS CVn stars listed in the \citet{2020ApJS..249...18C} catalogue, \citet{2021A&A...656A.125F} applied several restrictions based on known characteristics of ACV variables \citep[e.g.][]{2017MNRAS.468.2745N,2019A&A...622A.199J}: (a) variability period between one and ten days; (b) amplitude in the $r$ band of less than 0.3 mag; (c) the presence of a single independent variability frequency and corresponding harmonics; (d) stable or marginally changing light curve throughout the covered time span; and (e) an effective temperature between 6000 K and 25000 K \citep{andrae18}.

Item (c) was enforced to eliminate pulsating stars, which are mostly multi-periodic variables. Item (d) was chosen to distinguish ACV variables, whose light curves remain stable over very long periods of time (decades or more), from other spotted rotating variables, such as Sun-like stars or the aforementioned RS CVn variables, which are prone to exhibiting significant light changes as spots form and decay \citep[e.g.][]{kozhevnikova15}.

{Out of the 81\,393 stars identified as RS CVn stars by \citet{2020ApJS..249...18C},
1400 objects (1.7\%) were identified through these criteria, whose ZTF light curves were then downloaded
and visually inspected.} The most promising 86 ACV candidates were then selected and published by \citet{2021A&A...656A.125F}, who identified several photometric peculiarities of these stars that can be employed in the identification of mCP stars: (1) the light curves in $g$ and $r$ are in anti-phase; (2) the amplitude of variability is larger in $r$ than in $g$; (3) the light curves show inconsistent shapes in $g$ and $r$. Items (1) and (3) are, to the best of our knowledge, a peculiar characteristic of ACV variables and not seen in any other periodic variable stars. Item (2) is further proof that these stars are not pulsating variables, which are expected to show larger amplitudes at shorter wavelengths.

{A total of 82 objects were eliminated from the sample because they did not meet these criteria. So, the final sample consists of 1232 stars.} These objects were subsequently investigated in a CMD (cf. Sections \ref{CMD} and \ref{conclusion}); no RS CVn system, whose primary component is of spectral type F ($(BP-RP)_\mathrm{0}$\,$>$\,+0.5\,mag) or later, was found in this way.

{As a first step, the ZTF light curves of these 1232 stars were searched for periodicities using the packages Period04 \citep{2005CoAst.146...53L} 
and Peranso \citep{2016AN....337..239P}. 
Within the errors, we obtained the same results.}

\begin{table*}[t]
\caption{31 high-confidence ($P>0.7$) star cluster members cross-matched with the 
catalogue by \citet{hunt_2023}.}
\label{CPs_OCLs}
\begin{center}
\begin{tabular}{ccrccccccc}
\hline
\hline
ZTF ID & $G$ magnitude & Period & Host Cluster & $A_\mathrm{V}$ & e\_$A_\mathrm{V}$ & E\_$A_\mathrm{V}$ & $\log t$ & e\_$\log t$ & E\_$\log t$ \\
\hline
J200951.70+353418.1     &       13.831  &       0.521112        &       Biurakan\_2     &       1.24    &       1.17    &       1.31    &       6.85    &       6.84    &       6.90    \\
J062002.59+015217.6     &       13.797  &       2.379216        &       CWNU\_1313      &       1.13    &       0.98    &       1.23    &       8.85    &       8.28    &       8.88    \\
J033223.16+522840.4     &       14.495  &       1.928368        &       Czernik\_16     &       2.39    &       2.28    &       2.44    &       8.72    &       8.71    &       8.80    \\
J201432.04+411038.0     &       14.267  &       1.118963        &       FSR\_0219       &       2.82    &       2.63    &       2.96    &       6.80    &       6.78    &       6.90    \\
J215149.43+512546.7     &       14.326  &       1.969323        &       FSR\_0325       &       1.28    &       1.19    &       1.38    &       8.59    &       8.48    &       8.66    \\
J011734.53+611800.9     &       14.513  &       7.143945        &       FSR\_0534       &       2.24    &       2.08    &       2.40    &       7.82    &       7.70    &       8.09    \\
J044516.35+412701.6     &       14.200  &       25.026750       &       FSR\_0723       &       1.14    &       1.03    &       1.21    &       8.26    &       7.84    &       8.50    \\
J064008.37+132722.9     &       16.121  &       3.117477        &       FSR\_0975       &       1.13    &       0.97    &       1.35    &       8.70    &       8.41    &       8.77    \\
J052351.16+340638.2     &       13.659  &       6.944821        &       Gulliver\_53    &       1.24    &       1.15    &       1.34    &       8.18    &       8.11    &       8.49    \\
J043443.42+471535.0     &       15.452  &       2.195395        &       HSC\_1246       &       2.88    &       2.36    &       2.93    &       6.74    &       6.73    &       7.52    \\
J063607.12+171242.5     &       14.879  &       1.246997        &       HSC\_1552       &       2.15    &       2.04    &       2.37    &       8.42    &       7.85    &       8.70    \\
J070208.33-015427.8     &       13.817  &       2.943846        &       HSC\_1702       &       1.23    &       1.07    &       1.48    &       8.96    &       8.65    &       9.10    \\
J195926.03+300327.9     &       15.549  &       4.431191        &       HSC\_538        &       3.00    &       2.74    &       3.20    &       8.27    &       7.85    &       8.58    \\
J201440.55+344546.3     &       15.393  &       1.032562        &       HSC\_573        &       1.05    &       0.95    &       1.14    &       8.70    &       8.65    &       8.86    \\
J003024.62+592954.5     &       13.492  &       1.140750        &       HSC\_961        &       0.81    &       0.70    &       0.90    &       8.50    &       8.33    &       8.60    \\
J003241.55+614801.4     &       13.945  &       16.965972       &       King\_15        &       1.44    &       1.32    &       1.56    &       8.73    &       8.63    &       8.88    \\
J225208.24+581943.3     &       14.214  &       3.349355        &       King\_18        &       1.63    &       1.54    &       1.72    &       8.48    &       8.43    &       8.57    \\
J052008.16+392257.0     &       14.190  &       1.638088        &       NGC\_1857       &       1.90    &       1.65    &       1.95    &       7.59    &       7.48    &       7.76    \\
J014526.32+603634.2     &       14.472  &       3.436379        &       NGC\_659        &       1.54    &       1.52    &       1.57    &       7.30    &       7.23    &       7.32    \\
J195237.02+292309.4     &       13.820  &       1.855895        &       NGC\_6834       &       1.43    &       1.25    &       1.46    &       8.26    &       8.23    &       8.65    \\
J021932.01+570718.5     &       13.644  &       0.739862        &       NGC\_869        &       2.03    &       1.98    &       2.13    &       7.82    &       7.60    &       7.87    \\
J012949.33+623838.8     &       13.495  &       1.133402        &       OC\_0237        &       1.83    &       1.68    &       1.89    &       7.29    &       7.23    &       7.83    \\
J061911.26+183722.3     &       12.951  &       1.975130        &       Skiff\_J0619+18.5       &       0.74    &       0.63    &       0.84    &       8.41    &       8.12    &       8.60    \\
J002959.84+624914.9     &       15.164  &       1.362381        &       Theia\_1851     &       2.28    &       1.89    &       2.44    &       7.54    &       7.40    &       7.96    \\
J003005.72+625033.5     &       15.123  &       1.324113        &       Theia\_1851     &       2.28    &       1.89    &       2.44    &       7.54    &       7.40    &       7.96    \\
J200119.65+333657.0     &       14.636  &       2.926105        &       Toepler\_1      &       1.85    &       1.77    &       2.00    &       8.34    &       8.27    &       8.50    \\
J042455.95+441934.8     &       14.144  &       0.696192        &       UBC\_1263       &       2.18    &       1.97    &       2.36    &       7.51    &       7.41    &       7.59    \\
J042906.37+442429.1     &       14.275  &       1.029576        &       UBC\_1264       &       2.23    &       2.01    &       2.41    &       7.17    &       7.17    &       7.50    \\
J220008.02+461842.6     &       13.864  &       2.545526        &       UBC\_1600       &       1.66    &       1.55    &       1.77    &       8.30    &       8.10    &       8.49    \\
J061925.50+152011.7     &       15.945  &       2.171823        &       UBC\_438        &       1.58    &       1.47    &       1.64    &       8.69    &       8.64    &       8.71    \\
J201257.42+355238.9     &       14.645  &       0.925454        &       UBC\_580        &       0.48    &       0.39    &       0.56    &       8.84    &       8.70    &       8.87    \\

\hline
\end{tabular}
\end{center}
\end{table*}

\subsection{Classification} \label{classifications}

\subsubsection{Spectral classification} \label{mkclassification}

We used low-resolution LAMOST spectra from DR7, which contains almost 10 million archival spectra.\footnote{\url{http://www.lamost.org/dr7/v2.0/}} Only spectra with S/N\,$>$\,20 were considered. In summary, LAMOST spectra are available for { 138 out of 1232 stars (11.2\%)} 
stars of our sample.

Spectral classifications were derived with the modified MKCLASS code\footnote{MKCLASS is a computer program in C used to classify stellar spectra on the Morgan-Keenan-Kellman (MKK) spectral classification scheme \citep{2014AJ....147...80G}. The installation guide and further information can be accessed under \url{https://www.appstate.edu/~grayro/mkclass/}} (`MKCLASS\_mCP') of \citet{2020A&A...640A..40H}, which was specifically adapted to classify mCP stars, and the four libraries of standard star spectra used in the aforementioned study. For more information on the MKCLASS\_mCP code, its usage, and the standard libraries, we refer the reader to \citet{2020A&A...640A..40H}.

All spectral types derived in this way were sanity-checked manually. We stress that this process did not include a detailed manual classification on the  system, although in rare instances, weak peculiarities that were not identified by the code were manually added to the classifications. In the few instances where the code only provided the non-MKK-standard `bl4077' (strong blend at 4077\,\AA) and/or `bl4130' (strong blend at 4130\,\AA) classifications \citep[c.f.][]{2020A&A...640A..40H}, these were resolved into the contributing elements by visual inspection of the spectra. We caution, however, that the here derived peculiarity type classifications are not exhaustive as weak or complicated peculiarities may have been missed in several objects.

Furthermore, mCP stars often show narrow and peculiar hydrogen-line profiles that are easily misinterpreted as belonging to stars of higher luminosity. This also impacts automatic classification routines such as the MKCLASS code and is probably the reason for the high luminosity classification of many of our sample stars, which should be regarded with caution. It is well established that mCP stars are generally main-sequence objects \citep[e.g.][]{2017MNRAS.468.2745N}.

According to our expectations, all 138 stars turned out to be classical CP2 stars. The individual classifications are listed in Table \ref{table_master1}. Some representative example spectra are shown in Figure \ref{fig:showcase}.

      \begin{figure}
   \centering
   \includegraphics[width=\hsize]{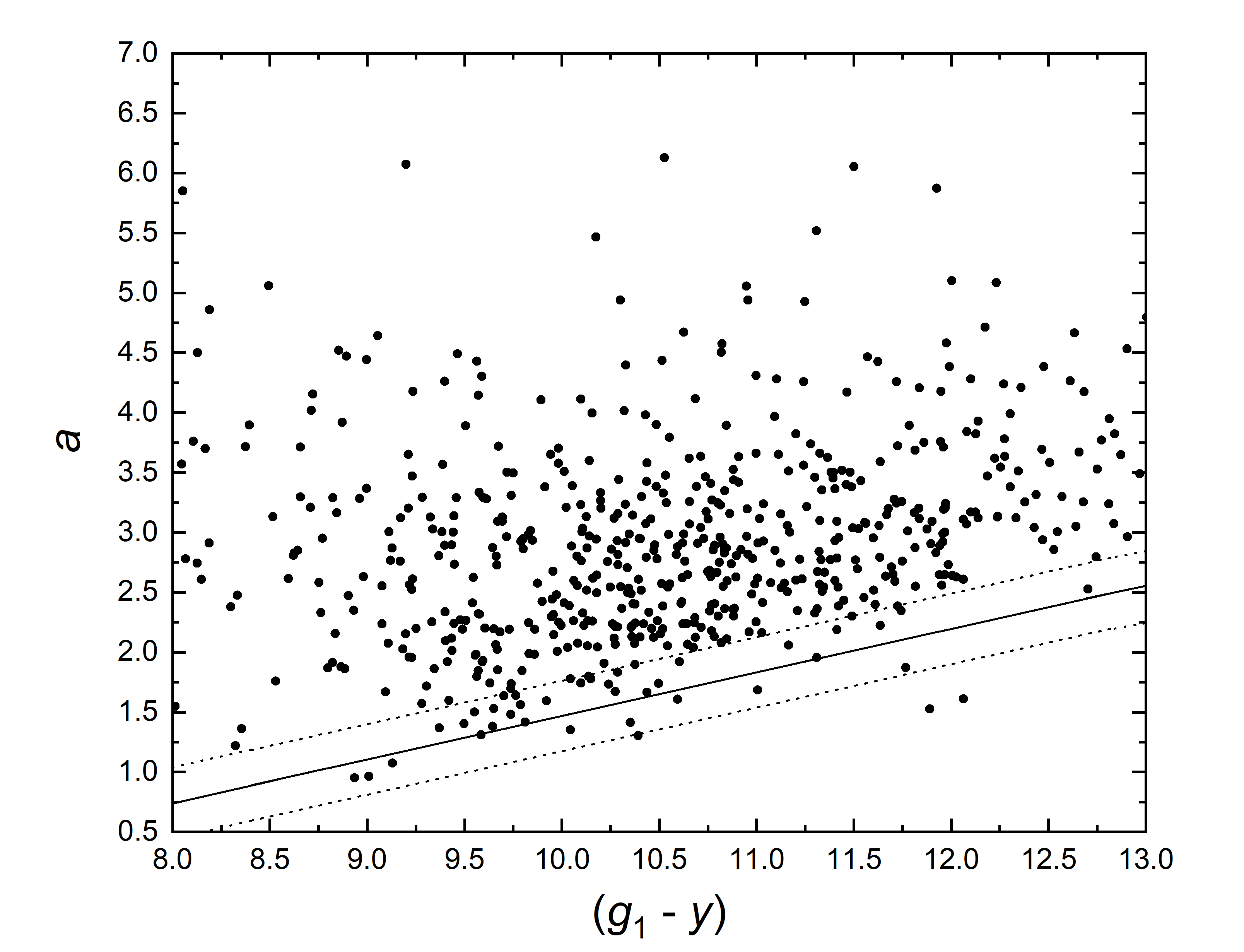}
      \caption{Calculated synthetic $\Delta$a magnitudes \citep{2022A&A...667L..10P} for the 665 sample stars with \textit{Gaia} BP/RP spectra. The dotted lines are
          the 95\% prediction bands used to select mCP stars.}
         \label{figure_Delta_a}
   \end{figure}

\subsubsection{Probing the 5200\,\AA\ flux depression}

Because LAMOST spectra are only available for a small fraction {(11.2\%)} of our sample stars, we also had to resort to alternative means of establishing chemical peculiarities. We therefore checked for the presence of a flux depression at around 5200\,\AA, a spectral feature that is characteristic of mCP stars and caused by line blanketing of Cr, Fe, and Si enhanced by a magnetic field \citep{2003MNRAS.341..849K,2007A&A...469.1083K}. While not all mCP stars show this flux depression, 
probably {because they have an unfavourable inclination} and the magnetic 
field characteristics \citep{2005A&A...441..631P}, probing this spectral region has become a staple in the search for new peculiar stars \citep[e.g.][]{maitzen80,maitzen98,2005A&A...441..631P,2020A&A...640A..40H,shi23}.

As a first step, the available LAMOST spectra (cf. Section \ref{mkclassification}) were visually checked for the presence of the 5200\,\AA\ flux depression. According to our expectations, of the 138 stars with LAMOST spectra, 132 stars clearly show this feature.

Another approach to utilising the 5200\,\AA\ flux depression was presented by \citet{2022A&A...667L..10P}. These authors used \textit{Gaia} BP/RP spectra \citep{carrasco21} for the synthesis of $\Delta$a photometry, which refers to a photometric system specifically developed to measure the 5200\,\AA\ flux depression \citep[e.g.][]{2005A&A...441..631P,stigler14}. They synthesised $\Delta$a photometry for 1240 known mCP stars from  BP/RP spectra and were able to effectively distinguish mCP stars from normal-type objects, with a detection level of more than 85\% for almost the entire investigated spectral range (almost 95\% for B- and A-type objects).

We followed the approach of \citet{2022A&A...667L..10P} and calculated synthetic $\Delta$a magnitudes for the 665 sample stars with \textit{Gaia} BP/RP spectra. In a nutshell, we normalised all spectra to the flux at 4020\,\AA\ and interpolated them in the wavelength region from 4800\,\AA\ to 5800\,\AA\ to a one-pixel resolution of 1\,\AA\ by applying a standard polynomial technique. Then the filter curves (with the central wavelengths as published by \citet{maitzen80b}: $g{_1}$ 5010\,\AA; $g{_2}$ 5215\,\AA; $y$ 5485\,\AA, and a bandwidth of 130\,\AA\ for each filter) were folded with the spectrum to calculate the synthetic magnitudes. Finally, all objects were investigated in the $a$ versus ($g1- y$) diagram. For more detailed information on this process, we refer the reader to the original paper \citep{2022A&A...667L..10P}.

According to the 3$\sigma$ level, 614 (92\%) of this subsample of 665 stars was detected as mCP stars (Fig. \ref{figure_Delta_a}). This is in accordance with the expectations from the abovementioned statistical studies about this group of stars. 

   \begin{figure*}
   \centering
   \includegraphics[width=0.9\hsize]{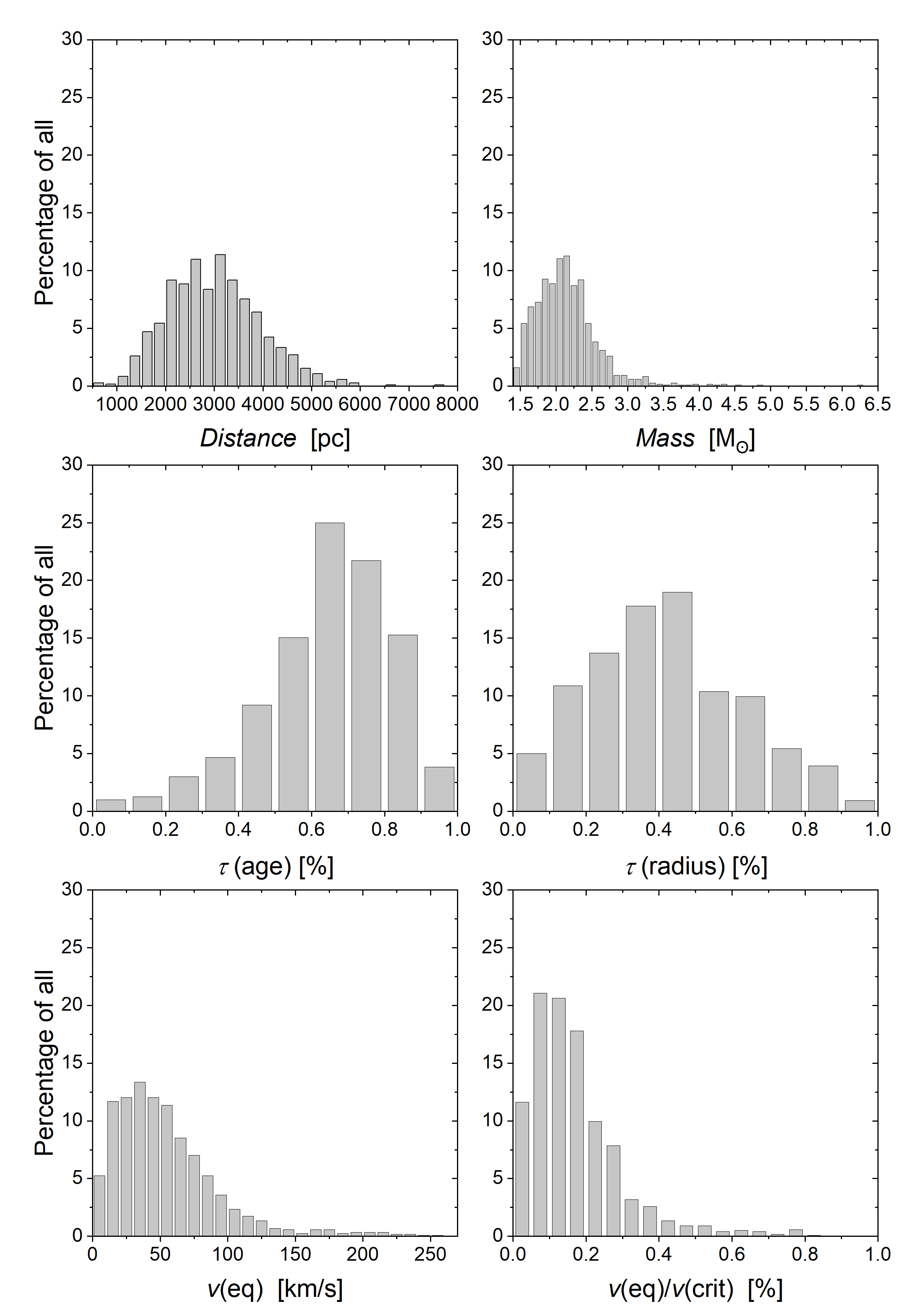}
      \caption{Distributions of the distance from the Sun (upper-left panel), mass (upper-right panel), fraction of life spent on the main sequence (middle-left panel), fraction of radius between the ZAMS and TAMS (middle-right panel), equatorial velocity (lower-left panel), and ratio of equatorial to critical velocity (lower-right panel).}
         \label{figure_histograms}
   \end{figure*}

\subsection{Colour-magnitude diagram, age, and radius} \label{CMD}

{The different \textit{Gaia} data releases do not list astrophysical parameters for all of
our target stars.}
Subsequently, the astrophysical properties of our sample stars were investigated in a CMD. To this end, we employed the homogeneous \textit{Gaia} DR2 photometry from \citet{gaia_dr2}.\footnote{We emphasise that the corresponding magnitudes from the more recent \textit{Gaia} DR3 do not change the overall results in a significant way.} For this photometric system, well-tested sets of PARSEC isochrones \citep{2012MNRAS.427..127B} are available.

All our sample stars are situated farther than 500\,pc from the Sun; therefore, interstellar reddening (absorption) cannot be neglected. To correct for this, we relied on the reddening map of \citet{green2019} and the photogeometric distances from
\citet{2021AJ....161..147B} to interpolate within this map. The transformation of the reddening values was performed using the relations\begin{equation}
E(B - V) = 0.76E(BP - RP) = 0.40A_\mathrm{G}.
\end{equation}
These relations already consider the conversion to extinction in different passbands using the coefficients listed in \citet{2018MNRAS.478..651G}. The absolute magnitude for the \textit{Gaia} $G$ magnitude ($M_{\rm G}$) of our target stars was calculated using the photogeometric distances and their errors from \citet{2021AJ....161..147B}.

Ages, masses, and radii were estimated by the Stellar Isochrone Fitting Tool\footnote{\url{https://github.com/Johaney-s/StIFT}}
\citep{2021osvm.confE...1S}, which follows the methods described by \citet{malkov10}. PARSEC isochrones \citep{2012MNRAS.427..127B} for a solar metallicity of Z\,=\,0.017 were used in this context, which has repeatedly been shown to be a good compromise for CP stars \citep[e.g.][]{2020A&A...640A..40H,2021A&A...645A..34P}.

To calculate the equatorial velocity ($v_\mathrm{equ}$) assuming rigid body rotation, we used the relation from \citet{1971PASP...83..571P}:
\begin{equation}
v_\mathrm{equ} = 50.79 R/P
,\end{equation}
where $R$ is the stellar radius in solar units, and $P$ is the observed period in days. To determine $\sigma v_\mathrm{equ}$, we only have to consider the
error of $R$ because $P$ is known with high precision. Recently, \citet{2023MNRAS.523.2440E} showed that $R$ can be determined with an error less than 3\% on a statistical basis. If we consider 5\% and 2M$_{\sun}$, we get $\sigma v_\mathrm{equ}$\,=\,5.1\,km\,s$^{-1}$. Subsequently, we calculated the first critical velocity ($v_\mathrm{crit}$) as adopted by
\citet{2014A&A...566A..21G} for the given metallicity (0.017) and individual stellar mass to obtain the velocity ratio $v_\mathrm{equ}$/$v_\mathrm{crit}$.

We extracted age and radius for a given mass at the terminal-age main sequence (TAMS), $t_\mathrm{TAMS}$ and $r_\mathrm{TAMS}$, from the 
PARSEC isochrone grid. Finally, the ratio $t/t_\mathrm{TAMS}$ and $r/r_\mathrm{TAMS}$ was calculated.

\section{Members of open clusters} \label{OCLs}

The membership of a studied star in a coeval star group, such as an open cluster, can be used to put tighter constraints
on the fundamental properties of the star, such as its distance, age, and initial metallicity, as all open cluster members share these properties
to a high degree of approximation. In such stellar groupings, the ensemble of several tens to hundred open cluster members allows the determination
of the properties of the studied star to significantly better precision and confidence, which would not be achievable if the star was studied in isolation. For this reason, identifying open cluster members among our studied sample of mCP stars might be of great value.

We used the star cluster membership lists from the catalogue of \citet{hunt_2023} to identify possible open cluster members among the studied stars.
This catalogue contains the parameters of 7\,167 star clusters, with more than 700 newly discovered high-confidence star clusters. In addition to this,
the study also includes a list of cluster members with membership probabilities for each listed star. We cross-matched our final sample with these lists and identified 31 high-confidence ($P>0.7$) star cluster members, which are presented in Table \ref{CPs_OCLs}.

In a subsequent step, we determined the reddening values and ages of the host open clusters. In contrast to a single star, through which multiple isochrones can pass, coeval stellar systems can lift the degeneracy by populating a large magnitude range in the CMD. In the limit of error-free measurements in conjunction with perfectly tuned isochrone models, determining precise stellar parameters is a trivial task achieved by minimising the geometric distances between the data and model curves. However, observational data may be influenced by factors like reddening, resolution limits, and contamination from the field. Other effects, such as intra-cluster metallicity variations or stellar rotation, can also cause deviations from model isochrones. Identifying the precise nature of each data point's deviation is often impossible, making it important to account for all these factors when choosing an appropriate statistical model.

We employed the isochrone fitting procedure from \citet{Ratzenbock2023b}, which models noise contributions around isochronal curves as samples drawn independently from skewed Cauchy distributions. The authors propose the skewed Cauchy distribution as it naturally incorporates non-symmetric noise sources, such as unresolved binaries and differential reddening effects inside the cluster. Additionally, the Cauchy distribution’s heavy tails are known for their robustness to outliers and ability to model data with many outliers \citep{Hampel2011}, decreasing the correlation strength of derived parameters to imperfections in the sample selection. Assuming a constant noise model for every data point along the isochrone, we obtain parameter uncertainties by applying Bayes’ theorem; the factorised Cauchy distribution becomes the likelihood while prior ranges are adopted from \citet{Ratzenbock2023b}.

We initialised Chronos by setting the metallicity to be constant (solar). Moreover, we used the distances from \citet{hunt_2023} and kept them fixed since they are expected to be already very precise. {They list mean errors of below 5\% for all their investigated clusters.} Extinction and age are kept as free parameters. Unlike in the case of using the reddening map of \citet{green2019}, we find that for clusters and Chronos, the best choice for transforming $A_\mathrm{V}$ to extinction in \textit{Gaia} bands is the following: $A_\mathrm{G}=0.857 A_\mathrm{V}$, $E(BP-RP)=0.518 A_\mathrm{V}$. These values were derived by applying the \citet{FitzpatrickMassa2007} extinction curves with $R_\mathrm{V}=3.1 - 3.3$ to the \textit{Gaia} DR3 passbands. The chosen extinction coefficients almost coincide with the constant terms of the polynomials recommended for \textit{Gaia} DR2 extinction, although they are somewhat shifted compared to the \textit{Gaia} DR3 polynomials.

For the first guess of cluster parameters in Chronos, we used the age and extinction values from \citet{hunt_2023} to set the corresponding prior ranges to be within one order of magnitude in logarithmic age and within 1.5\,mag around these values, respectively. The initial fit was then used to manually set the ranges (usually decreasing them), which were then used in the statistical fitting performed by Chronos. The lower and upper bounds of the extracted parameters are at 5 and 95 percentiles, respectively. All results are presented in Table \ref{CPs_OCLs}.

It should be pointed out that some of the clusters show a complicated posterior distribution. This is caused either by having a small sample of stars that do not populate a main sequence or by having many outliers in the red giant regions. For this reason, Chronos was not able to find a narrow range of cluster parameters for the following objects: CWNU~1313, FSR~0975, HSC~538, HSC~1552, King~15, NGC~1857, and UBC~1264. Moreover, the procedure did not correctly fit the parts of the main sequences of NGC~659 and UBC~1264, which are least affected by differential reddening. Finally, we note that the data distribution in the CMD of UBC~1264 shows an uncharacteristically prominent knee (at $BP-RP=0.5$ mag) for a cluster of the fitted age. However, the isochrone corresponding to the lower bounds of age and extinction fit the data very well.

We have identified three stars that are members of star clusters younger than 10\,Myr. According to their locations in the CMD, these objects just reached the zero-age main sequence (ZAMS). Then, there is a continuous sequence until the age of about 1\,Gyr. This supports the conclusions from the field star sample, which are described in the following section.

   \begin{figure}
   \centering
   \includegraphics[width=\hsize]{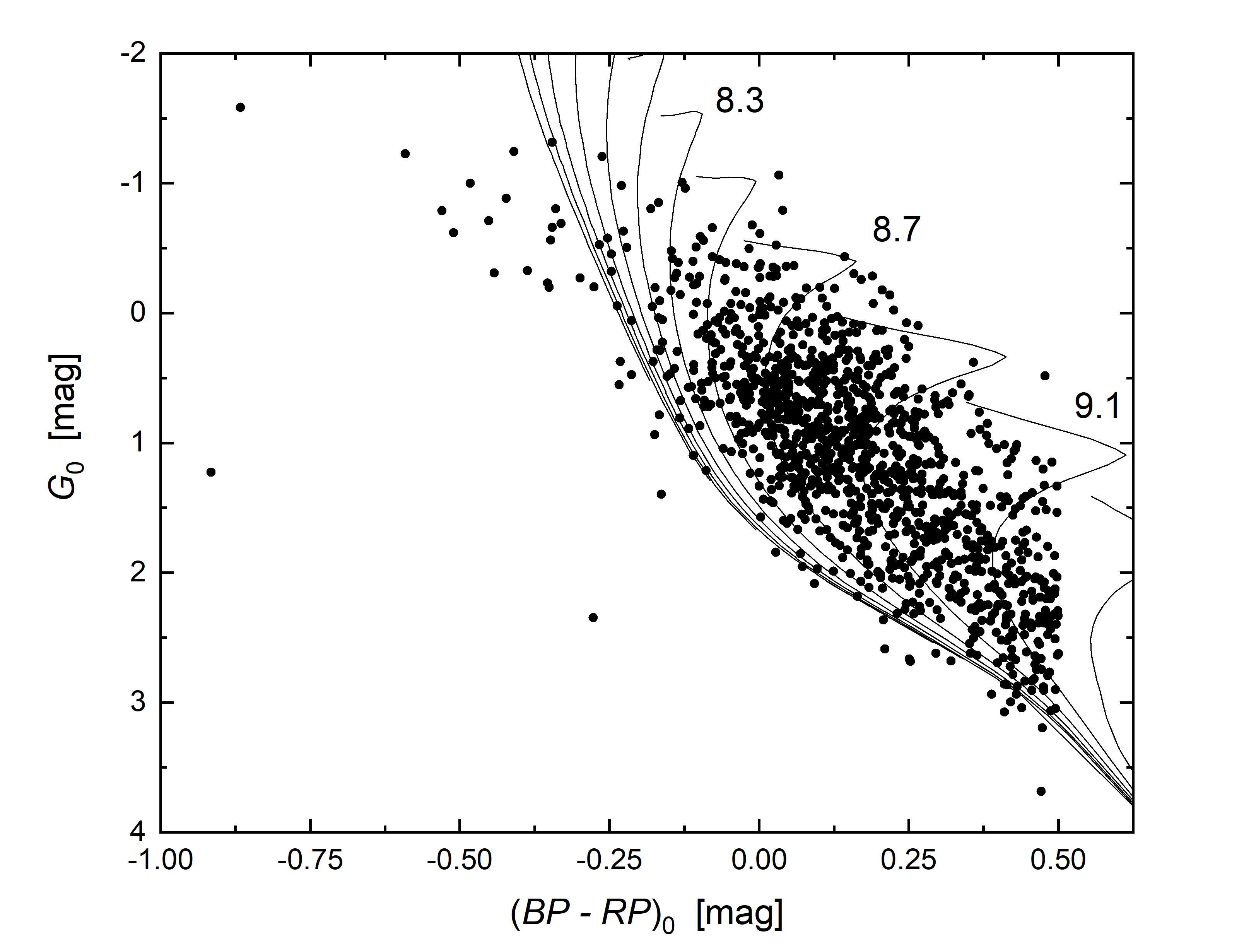}
      \caption{$G_0$ versus $(BP-RP)_0$ diagram for the target star sample together with isochrones from 
      \citet{2012MNRAS.427..127B} for solar metallicity [Z]\,=\,0.017\,dex. {The values refer to the logarithmic ages of the isochrones.} There is an obvious lack of stars
      close to the ZAMS.}
         \label{figure_CMD}
   \end{figure}
   
\section{Discussion} \label{discussion}

This section summarises the results from our analysis, discussing different astrophysical aspects. Histograms illustrating the distributions of the distance from the Sun, mass, fraction of life spent on the main sequence, fraction of radius between ZAMS and TAMS, equatorial velocity, and ratio of equatorial to critical velocity are provided in Figure \ref{figure_histograms}.

{\it CMD:} Figure \ref{figure_CMD} presents the $G_\mathrm{0}$ versus $(BP-RP)_\mathrm{0}$ diagram for the target star 
sample together with isochrones from \citet{2012MNRAS.427..127B} for solar metallicity Z\,=\,0.017\,dex.
We find 35 objects (3\%) not covered by the isochrone grid, in accordance with the results from other studies \citep{2020A&A...640A..40H}. The reasons are unknown. {The preference away from the ZAMS is apparent visually and cannot be explained with the chosen metallicity \citep{2021A&A...656A.125F}.}

{\it Distance:} Most of our sample stars are situated between 1000 and 6000\,pc from the Sun. They are populating the
first three quadrants in the Galactic disk, where the reddening cannot be neglected. All stars are members
of the thin or thick disk \citep{2022ApJ...932...28V}.

{\it Mass:} We see a peak of the mass distribution at about two solar masses, which corresponds to a spectral type of about A2. Otherwise, the spectral range from F1 to B8 is well populated with an apparent lack of more massive stars. The reason is unclear: there was no selection effect on astrophysical parameters. 

{\it $\tau$(age)} and {\it $\tau$(radius):} It is clear that mCP stars are already present when they reach the ZAMS, as reported
before in studies that investigated members of young open clusters \citep{2005A&A...441.1111P}. Subsequently, a constant increase in the number of stars up to a peak at about 65\% of the main-sequence lifetime is reached. The diffusion timescale is much shorter and therefore not adequate to account for these observations \citep{michaud1981}. One explanation might be an active dynamo that is driving and enhancing the magnetic field. However, for such a dynamo to work, differential rotation is needed \citep{2011MNRAS.411.1059K}, which is only observed for a small
fraction of upper main-sequence \citep{2004A&A...415..325R} and mCP \citep{2018CoSka..48..203M} stars. 
It was stated by \citet{2012A&A...542A.116A} that no differential rotators are known with effective temperatures higher than 7400\,K, a fact that has not been challenged since.

The distribution of the relative radii of our sample stars is quite different. For a constant mass, the radius increases
from ZAMS to TAMS by a factor of about 2.4. Assuming that the total magnetic flux is conserved,
it is distributed over a larger surface at the TAMS and thus becomes locally weaker. This influences the 
diffusion rates and the overall element abundances on the surface. The $\tau$(radius) distribution is symmetric with a
peak at 45\% of the TAMS value and therefore different than the distribution of ages. Together, these two distributions define the astrophysical parameter space where mCP stars are most commonly found. This is essential information not only for models but also for a further search for new class members.

{\it Rotational velocity:} Models predict that surface abundance patterns created by the diffusion of elements are destroyed by mixing induced by meridional circulation as a consequence of rotation \citep{2003ASPC..305..199T}. However, the
actual critical rotational velocity depends on the stellar mass and the strength of the magnetic field \citep{1976ApJ...209..816S}.
The guide value is about 125\,km\,s$^{-1}$. This amounts to about 40\% of the critical velocity. In total, we identified 38 stars exceeding a $v_\mathrm{equ}$ of 150\,km\,s$^{-1}$, with extreme values up to 260\,km\,s$^{-1}$. The estimated statistical error is about 5\,km\,s$^{-1}$.

We double-checked the 38 fast rotators to probe for any light curve or other peculiarities that may hint that these stars are not ACV variables and hence classical mCP stars. However, none were found: all stars show light variability in accordance with rotational modulation and pass the strict criteria imposed in the selection process (cf. Section \ref{target_selection}).

A possible explanation is that we see the fast rotation of a cool companion star in an
undetected spectroscopic binary system. It is well known that such short-period objects exist 
\citep{2022ApJ...934..146S,2022AJ....164..251L}. However, such a possible companion would be between six to ten magnitudes fainter than
the primary component. We can therefore rule out this scenario if we scale the observed amplitude to those surveyed in the listed references.

\section{Conclusion} \label{conclusion}

The available spectral data confirm that the employed selection process is highly efficient for detecting ACV variables and that the final sample of 1232 stars can be regarded as a pure sample of mCP stars.

In summary, we are confident that all 38 fast rotating stars are ACV variables. This interesting finding is in line with the results of \citet{mikulasek22}, who identified an equatorial velocity of $v_\mathrm{equ}$ = 230\,km\,s$^{-1}$ for the well-established CP2 star HD\,60431 and should now be aligned with the oblique rotator and diffusion models. Basic questions, such as how surface spots can manifest at such high rotational velocities, are still unanswered. Maybe a strong magnetic field stabilises the stellar atmosphere sufficiently. We therefore strongly encourage further detailed studies of these objects, in particular polarimetric measurements or the search for Zeeman splitting of lines, which will shed more light on the individual magnetic field characteristics. Due to the relative faintness of our target stars, however, such measurements can only be performed with the largest ground-based telescopes and a significant amount of observing time.
   
\begin{acknowledgements}
We want to thank the referee for carefully reading our manuscript and giving such constructive comments, which substantially helped improve the quality of the paper.
M.~Pri{\v s}egen is supported by the European Regional Development Fund, project No. ITMS2014+: 313011W085.
Based on observations obtained with the Samuel Oschin 48-inch Telescope at the Palomar Observatory as part of the \textit{Zwicky} Transient Facility project, which is supported by the National Science Foundation under Grant No. AST-1440341 and a collaboration including Caltech, IPAC, the Weizmann Institute for Science, the Oskar Klein Center at Stockholm University, the University of Maryland, the University of Washington, Deutsches Elektronen-Synchrotron and Humboldt University, Los Alamos National Laboratories, the TANGO Consortium of Taiwan, the University of Wisconsin at Milwaukee, and Lawrence Berkeley National Laboratories. Operations are conducted by COO, IPAC, and UW. Guoshoujing Telescope (the Large Sky Area Multi-Object Fiber Spectroscopic Telescope LAMOST) is a National Major Scientific Project built by the Chinese Academy of Sciences. Funding for the project has been provided by the National Development and Reform Commission. LAMOST is operated and managed by the National Astronomical Observatories, Chinese Academy of Sciences. This work has made use of data from the European Space Agency (ESA) mission {\it Gaia} (\url{https://www.cosmos.esa.int/gaia}), processed by the {\it Gaia} Data Processing and Analysis Consortium (DPAC, \url{https://www.cosmos.esa.int/web/gaia/dpac/consortium}). Funding for the DPAC has been provided by national institutions, in particular the institutions
participating in the {\it Gaia} Multilateral Agreement.
This publication makes use of data products from the Two Micron All Sky Survey, which is a joint project of the University of Massachusetts and the Infrared Processing and Analysis Center/California Institute of Technology, funded by the National Aeronautics and Space Administration and the National Science Foundation.

\end{acknowledgements}

\bibliographystyle{aa} 
\bibliography{paper_bibliography.bib}

\begin{appendix} 

\section{Essential data and light curves for our sample stars}

Table \ref{table_master1},
available in full at the CDS, lists the essential data for our sample stars.

\clearpage

\setcounter{table}{0}  
\begin{table*}
\caption{Essential data for our sample stars, sorted by increasing right ascension (extract). The full table is available at the CDS.\ The columns denote: 
(1) ZTF ID; (2) ID; (3) Right ascension (J2000; \textit{Gaia} EDR3); (4) Declination (J2000; $aia$ EDR3); 
(5) Period from \citet{2020ApJS..249...18C}; (6) $G$ magnitude (\textit{Gaia} DR2); (7) $G$ magnitude error (\textit{Gaia} DR2);
(8) $(BP-RP)$ colour (\textit{Gaia} DR2); (9) $(BP-RP)$ colour error;
(10) $(g1-y)$ colour; (11) $a$ index;
(12) $E(B-V)$; (13) Parallax (\textit{Gaia} DR3);
(14) Parallax error; (15) Mass; (16) Fractional age on the main sequence;
(17) Fractional radius on the main sequence;
(18) Equatorial velocity; (19) Ratio of the equatorial and breakup velocity;
(20) Spectral type.}  
\label{table_master1}
\begin{center}
\begin{adjustbox}{max width=1.2\textwidth,angle=90}
\begin{tabular}{lllcclccccccccccccccccccc}
\hline
\hline
(1) & (2) & (3) & (4) & (5) & (6) & (7) & (8) & (9) & (10) & (11) & (12) & (13) & (14) & (15)  & (16)  & (17)  & (18)  & (19)  & (20)  \\
\hline
ZTFJ000041.86+631059.3  &       430083701322198016      &       0.174456        &       63.183145       &       2.022767        &       13.3856 &       0.0009  &       0.4354  &       0.0041  &       9.741   &       1.737   &       0.341   &       0.3222  &       0.0117  &       2.53    &       0.76    &       0.53    &       78.3    &       0.24    &               \\
ZTFJ000049.14+580009.1  &       422632585891909632      &       0.204771        &       58.002501       &       4.233268        &       13.0044 &       0.0012  &       0.7462  &       0.0050  &       11.719  &       3.243   &       0.408   &       0.5623  &       0.0157  &       2.12    &       0.79    &       0.62    &       36.6    &       0.12    &               \\
ZTFJ000049.29+560741.1  &       420909067055294976      &       0.205423        &       56.128087       &       3.187368        &       14.6264 &       0.0011  &       0.6924  &       0.0058  &       10.439  &       1.664   &       0.200   &       0.3366  &       0.0219  &       1.67    &       0.69    &       0.42    &       35.6    &       0.12    &               \\
ZTFJ000107.80+630653.9  &       430082911048271360      &       0.282551        &       63.114962       &       1.451780        &       13.2359 &       0.0008  &       0.5658  &       0.0042  &       10.676  &       2.041   &       0.410   &       0.3155  &       0.0105  &       2.65    &       0.88    &       0.80    &       136.0   &       0.46    &               \\
ZTFJ000109.59+632305.7  &       431593640023939200      &       0.289977        &       63.384897       &       4.230083        &       13.8605 &       0.0008  &       0.6201  &       0.0039  &       10.337  &       2.706   &       0.415   &       0.2883  &       0.0129  &       2.38    &       0.80    &       0.65    &       39.7    &       0.13    &               \\
ZTFJ000310.20+631736.2  &       431576975551028736      &       0.792539        &       63.293393       &       3.651281        &       15.7526 &       0.0010  &       1.1375  &       0.0056  &               &               &       0.578   &       0.1761  &       0.0286  &       2.00    &       0.85    &       0.79    &       46.1    &       0.17    &               \\
ZTFJ000318.72+592646.6  &       423269993399745408      &       0.828042        &       59.446267       &       1.889031        &       14.9951 &       0.0015  &       0.9864  &       0.0062  &       9.614   &       3.280   &       0.540   &       0.3543  &       0.0215  &       1.83    &       0.64    &       0.38    &       61.5    &       0.19    &               \\
ZTFJ000329.10+595719.9  &       423313355390307200      &       0.871295        &       59.955527       &       2.194480        &       13.1024 &       0.0010  &       0.3591  &       0.0050  &       11.126  &       3.154   &       0.633   &       0.3212  &       0.0112  &               &               &               &               &               &               \\
ZTFJ000340.89+644505.5  &       432139513181384192      &       0.920394        &       64.751531       &       8.445032        &       14.0842 &       0.0012  &       1.0062  &       0.0062  &       13.005  &       4.796   &       0.703   &       0.2875  &       0.0168  &       2.58    &       0.93    &       0.97    &       25.6    &       0.09    &               \\
ZTFJ000340.91+623146.9  &       430027179553703552      &       0.920462        &       62.529681       &       2.005254        &       14.7932 &       0.0031  &       0.8354  &       0.0138  &               &               &       0.563   &       0.2510  &       0.0198  &       2.22    &       0.74    &       0.53    &       73.4    &       0.24    &               \\
ZTFJ000400.12+642128.1  &       431749564514884352      &       1.000544        &       64.357803       &       2.433503        &       16.6064 &       0.0013  &       1.2179  &       0.0085  &               &               &       0.656   &       0.1907  &       0.0435  &       1.72    &       0.64    &       0.36    &       45.6    &       0.14    &               \\
ZTFJ000400.34+634528.6  &       431617039007033472      &       1.001421        &       63.757946       &       1.722651        &       16.3320 &       0.0019  &       0.8074  &       0.0098  &               &               &       0.487   &       0.2220  &       0.0375  &       1.76    &       0.19    &       0.05    &       47.1    &       0.13    &               \\
ZTFJ000430.91+633115.1  &       431607212121962112      &       1.128801        &       63.520859       &       2.814172        &       16.3703 &       0.0019  &       1.0169  &       0.0087  &               &               &       0.471   &       0.1778  &       0.0417  &       1.69    &       0.67    &       0.39    &       39.7    &       0.13    &               \\
ZTFJ000453.11+643446.7  &       431759391399934464      &       1.221324        &       64.579654       &       4.599917        &       15.7751 &       0.0006  &       1.1203  &       0.0041  &       10.804  &       2.301   &       0.538   &       0.4689  &       0.0587  &       1.49    &       0.13    &       0.02    &       16.1    &       0.04    &               \\
ZTFJ000500.54+640951.9  &       431723794710271488      &       1.252267        &       64.164416       &       4.946422        &       14.2421 &       0.0012  &       0.8109  &       0.0055  &       13.323  &       3.444   &       0.775   &       0.1991  &       0.0152  &       3.82    &       0.76    &       0.48    &       38.0    &       0.11    &               \\
ZTFJ000520.70+604135.4  &       429317960188135296      &       1.336262        &       60.693161       &       5.213785        &       12.6493 &       0.0008  &       0.6882  &       0.0052  &               &               &       0.319   &       0.9322  &       0.0109  &       1.79    &       0.57    &       0.29    &       20.3    &       0.06    &               \\
ZTFJ000526.28+583002.9  &       423076440687524992      &       1.359555        &       58.500809       &       2.186827        &       15.5857 &       0.0013  &       0.9696  &       0.0056  &       12.774  &       3.770   &       0.605   &       0.1530  &       0.0288  &       2.24    &       0.83    &       0.71    &       77.9    &       0.27    &               \\
ZTFJ000527.78+625005.0  &       430014225928748928      &       1.365792        &       62.834728       &       4.349762        &       14.3635 &       0.0007  &       0.6456  &       0.0038  &               &               &       0.558   &       0.2277  &       0.0163  &       2.78    &       0.73    &       0.47    &       36.4    &       0.11    &               \\
ZTFJ000551.03+624510.1  &       430012645380856960      &       1.462665        &       62.752815       &       3.671449        &       15.1011 &       0.0009  &       0.7980  &       0.0043  &               &               &       0.644   &       0.2391  &       0.0212  &       2.42    &       0.59    &       0.31    &       34.7    &       0.10    &               \\
ZTFJ000618.06+623707.3  &       430007319621536256      &       1.575269        &       62.618700       &       3.289776        &       15.9185 &       0.0009  &       1.0531  &       0.0051  &               &               &       0.628   &       0.2158  &       0.0301  &       1.90    &       0.64    &       0.38    &       36.2    &       0.11    &               \\
ZTFJ000635.95+594051.6  &       423247453410415744      &       1.649819        &       59.680993       &       1.659816        &       15.2871 &       0.0010  &       0.8731  &       0.0064  &               &               &       0.535   &       0.2778  &       0.0248  &       1.91    &       0.57    &       0.29    &       66.2    &       0.20    &               \\
ZTFJ000713.79+634446.5  &       431660465409284992      &       1.807494        &       63.746253       &       1.362058        &       15.2988 &       0.0008  &       0.5878  &       0.0043  &               &               &       0.395   &       0.2418  &       0.0231  &       2.00    &       0.40    &       0.16    &       72.1    &       0.20    &               \\
ZTFJ000714.47+670842.6  &       528526997434410880      &       1.810308        &       67.145144       &       5.153270        &       16.2636 &       0.0019  &       2.3053  &       0.0115  &       10.126  &       3.132   &       1.610   &       0.4019  &       0.0378  &       2.23    &       0.85    &       0.79    &       34.7    &       0.12    &               \\
ZTFJ000728.72+604238.5  &       429397571702143616      &       1.869691        &       60.710681       &       4.813121        &       14.7596 &       0.0005  &       0.6558  &       0.0030  &       14.459  &       4.126   &       0.495   &       0.2336  &       0.0223  &       2.33    &       0.66    &       0.39    &       28.0    &       0.08    &               \\
ZTFJ000745.46+603122.5  &       429300608520427008      &       1.939435        &       60.522902       &       6.683301        &       14.7356 &       0.0003  &       0.8876  &       0.0034  &       8.054   &       5.848   &       0.366   &       0.5060  &       0.0189  &       1.55    &       0.46    &       0.17    &       13.0    &       0.04    &               \\
ZTFJ000858.22+694045.7  &       530739146110430080      &       2.242614        &       69.679361       &       2.926237        &       13.4953 &       0.0012  &       1.4276  &       0.0051  &       11.945  &       3.756   &       0.920   &       0.5995  &       0.0109  &       2.22    &       0.98    &       1.05    &       71.9    &       0.28    &               \\
ZTFJ001019.45+644548.8  &       432102267224418176      &       2.581093        &       64.763570       &       4.087818        &       13.8197 &       0.0014  &       0.8993  &       0.0063  &       10.654  &       3.617   &       0.625   &       0.4280  &       0.0126  &       2.28    &       0.76    &       0.55    &       37.3    &       0.12    &               \\
ZTFJ001038.63+594220.8  &       429059025197785728      &       2.660965        &       59.705776       &       6.943025        &       14.6724 &       0.0006  &       1.0211  &       0.0063  &               &               &       0.507   &       0.4086  &       0.0206  &       1.79    &       0.71    &       0.46    &       17.7    &       0.06    &               \\
ZTFJ001056.46+642626.5  &       431887072184295424      &       2.735301        &       64.440717       &       2.739042        &       15.5147 &       0.0010  &       1.1315  &       0.0051  &       9.984   &       2.249   &       0.898   &       0.1976  &       0.0280  &       2.60    &       0.75    &       0.53    &       58.6    &       0.18    &               \\
ZTFJ001112.69+574452.5  &       422472572590872960      &       2.802903        &       57.747908       &       0.765706        &       12.7760 &       0.0006  &       0.4041  &       0.0035  &               &               &       0.273   &       0.6490  &       0.0134  &       2.09    &       0.47    &       0.21    &       138.6   &       0.39    &               \\
ZTFJ001153.24+643005.5  &       431907275711549824      &       2.971880        &       64.501523       &       12.892051       &       15.9590 &       0.0014  &       1.0598  &       0.0061  &       10.950  &       2.967   &       0.884   &       0.1669  &       0.0326  &       2.61    &       0.56    &       0.28    &       9.9     &       0.03    &               \\
ZTFJ001204.14+623936.6  &       431310378339767168      &       3.017297        &       62.660170       &       1.808221        &       13.6627 &       0.0007  &       0.4666  &       0.0033  &       12.271  &       4.238   &       0.356   &       0.3780  &       0.0130  &       2.30    &       0.59    &       0.32    &       68.8    &       0.20    &               \\
ZTFJ001218.47+603439.1  &       429218729264606592      &       3.076986        &       60.577530       &       6.995805        &       14.1988 &       0.0007  &       0.9023  &       0.0051  &       10.269  &       2.117   &       0.517   &       0.3211  &       0.0157  &       2.20    &       0.84    &       0.75    &       24.7    &       0.09    &               \\
ZTFJ001322.71+633434.6  &       431470632159430144      &       3.344665        &       63.576282       &       2.437494        &       13.1437 &       0.0008  &       0.4907  &       0.0050  &               &               &       0.428   &       0.3060  &       0.0122  &       2.90    &       0.83    &       0.65    &       76.7    &       0.24    &               \\
ZTFJ001339.71+602722.9  &       429167288428957568      &       3.415491        &       60.456366       &       1.533054        &       16.1140 &       0.0020  &       1.0177  &       0.0133  &               &               &       0.629   &       0.2463  &       0.0367  &       1.81    &       0.46    &       0.19    &       63.3    &       0.18    &               \\
ZTFJ001445.88+623546.0  &       431254028366483968      &       3.691193        &       62.596110       &       1.909181        &       13.7831 &       0.0007  &       0.5602  &       0.0051  &               &               &       0.696   &       0.2483  &       0.0140  &               &               &               &               &               &               \\
ZTFJ001446.92+625710.7  &       431287769630500480      &       3.695520        &       62.952972       &       2.299285        &       16.3118 &       0.0024  &       1.0965  &       0.0112  &               &               &       0.609   &       0.2276  &       0.0368  &       1.74    &       0.58    &       0.29    &       45.6    &       0.14    &               \\
ZTFJ001450.13+631219.8  &       431433145683973888      &       3.708931        &       63.205505       &       1.270681        &       15.2653 &       0.0010  &       0.8220  &       0.0060  &       12.634  &       4.664   &       0.514   &       0.2960  &       0.0234  &       1.90    &       0.48    &       0.21    &       79.6    &       0.23    &               \\
ZTFJ001457.48+623501.0  &       431253856567801088      &       3.739507        &       62.583618       &       3.127105        &       16.3299 &       0.0010  &       1.4402  &       0.0069  &               &               &       0.763   &       0.1597  &       0.0399  &       1.92    &       0.89    &       0.88    &       55.3    &       0.21    &               \\
ZTFJ001519.54+634008.1  &       431428300959455744      &       3.831463        &       63.668919       &       15.320009       &       14.4333 &       0.0010  &       0.9377  &       0.0049  &       14.053  &       4.649   &       0.634   &       0.2123  &       0.0162  &       2.47    &       0.99    &       1.10    &       15.0    &       0.06    &               \\
ZTFJ001542.54+624753.9  &       431279832531026048      &       3.927311        &       62.798316       &       3.139875        &       15.5410 &       0.0010  &       1.0720  &       0.0058  &       10.235  &       2.756   &       0.621   &       0.2058  &       0.0257  &       2.07    &       0.80    &       0.65    &       49.6    &       0.17    &               \\
ZTFJ001545.14+643417.7  &       431946789406738432      &       3.938129        &       64.571592       &       1.782231        &       14.7195 &       0.0008  &       0.9810  &       0.0038  &       10.179  &       2.493   &       0.687   &       0.3476  &       0.0188  &       2.16    &       0.67    &       0.42    &       74.2    &       0.23    &               \\
ZTFJ001601.82+605057.1  &       429545627813160192      &       4.007604        &       60.849202       &       5.273914        &       14.8017 &       0.0010  &       0.9200  &       0.0060  &       10.376  &       2.402   &       0.515   &       0.3218  &       0.0193  &       1.96    &       0.70    &       0.48    &       25.1    &       0.08    &               \\
ZTFJ001619.48+584544.4  &       422955292553097472      &       4.081181        &       58.762330       &       1.636320        &       14.8918 &       0.0006  &       0.8969  &       0.0034  &       10.991  &       2.569   &       0.499   &       0.2812  &       0.0191  &       2.01    &       0.75    &       0.54    &       85.9    &       0.28    &               \\
ZTFJ001639.31+632910.3  &       431417134044654080      &       4.163834        &       63.486206       &       2.958571        &       13.9559 &       0.0010  &       0.6424  &       0.0048  &               &               &       0.490   &       0.2376  &       0.0155  &       2.69    &       0.86    &       0.75    &       65.3    &       0.22    &               \\
ZTFJ001648.83+624310.2  &       431281310000032128      &       4.203495        &       62.719501       &       3.106824        &       13.7082 &       0.0009  &       0.5887  &       0.0043  &       11.992  &       4.385   &       0.391   &       0.4080  &       0.0124  &       2.15    &       0.63    &       0.38    &       41.0    &       0.12    &               \\
ZTFJ001717.85+614459.2  &       429711997667191936      &       4.324396        &       61.749780       &       2.191678        &       15.7613 &       0.0012  &       1.1243  &       0.0064  &       10.715  &       3.633   &       0.685   &       0.2655  &       0.0301  &       1.91    &       0.66    &       0.40    &       55.3    &       0.17    &               \\
ZTFJ001742.73+583243.8  &       422200237304252800      &       4.428095        &       58.545489       &       1.864920        &       15.0093 &       0.0008  &       0.9108  &       0.0052  &       10.751  &       2.674   &       0.433   &       0.3467  &       0.0202  &       1.73    &       0.62    &       0.34    &       58.5    &       0.18    &               \\
ZTFJ001751.41+593607.4  &       428263700334022528      &       4.464262        &       59.602048       &       1.289917        &       14.2339 &       0.0008  &       0.7667  &       0.0029  &       11.420  &       2.543   &       0.529   &       0.4321  &       0.0160  &       2.05    &       0.52    &       0.25    &       84.7    &       0.24    &               \\
ZTFJ001843.03+725256.6  &       537329412648335616      &       4.679357        &       72.882386       &       2.414676        &       13.1813 &       0.0015  &       0.4760  &       0.0075  &               &               &       0.330   &       0.3535  &       0.0106  &       2.46    &       0.81    &       0.64    &       70.4    &       0.23    &               \\
\hline
\end{tabular}  
\end{adjustbox}  
\end{center}   
\end{table*}

\end{appendix}

\end{document}